# Nd:YAG single crystals grown by floating zone method in laser furnace

František Zajíc, Martin Klejch, Adam Eliáš, Milan Klicpera, Alena Beitlerová, Martin Nikl, and Jiří Pospíšil*

ABSTRACT: We report on single crystal growth of laser material Nd:YAG widely used in the applications by innovative crucible-free floating zone method implemented in an advanced laser optical furnace. We have optimized the parameters for the production of high-quality single crystals of the size typical for laser tubes. To reduce the strain and improve machinability, we have developed an afterheater to thermalize the grown part of a single crystal below the hot zone, which is a technique unavailable in common mirror furnaces. The high quality of the single crystals was verified by Laue diffraction and the internal strain was documented by neutron diffraction. The absorption spectrum corresponds with the parameters of the commercially used material produced by Czochralski method. The presented methodology for the single crystal growth by floating zone method with laser heating is applicable for the preparation of other high-quality single crystals of oxide-based materials, particularly in an oxidizing environment unattainable in commonly used crucible methods.

KEYWORDS: Floating zone method, laser furnace, Nd:YAG, single crystal,

## INTRODUCTION

The demand for materials in the form of single crystals for various applications has a growing trend. The customer's requests are focused on the quantization of the grown amounts of single crystals, reproducibility in production to reduce very expensive single-crystalline scrap, and, particularly, the improvement of their functionalities to enlarge the utilization. All these aspects are solved by a variety of growth techniques, which are fit to specific thermodynamic properties of each material combined with the technological aspects to produce single crystals of appropriate shape, size, and quality. The majority amount of nowadays single crystals is grown by Verneuil [1], Bridgman–Stockbarger, temperature-gradient technique [2, 3], and Czochralski methods [4]. Namely, the last one has passed by technological progress [4] thanks to its utilization in the production of silicon single crystals or optical materials in the field of scintillators [5-7] and solid-state lasers [8-11]. However, the substantial limitation of the Czochralski method is the crucible which is a weak point of this process in several aspects [12, 13]. The chemically aggressive high-temperature melts are often contaminated by the material of gradually dissolved crucibles. It is well documented in the case of silicon single crystals grown from quartz crucibles [14]. The crucible obstacles are even more serious in the case of high-temperature melting scintillators where very expensive noble metals Rh or Ir crucibles are necessary to use. Moreover, the metallic materials are very sensitive to residual oxygen at high temperatures which reserve the growth processes only in an inert or reducing environment although the work in pure oxygen under the pressure can significantly improve the optical properties of the oxide-based single crystals [15].

Our work primarily focuses on single crystal growth of scintillators and solid-state lasers based on a garnet structure derived from yttrium-aluminum garnet (YAG). The substituted variants of laser materials Nd:YAG [16], Er:YAG [17], Yb:YAG [18], Tm:YAG [19] or scintillator (Ce:YAG [20]) have found very broad applications. The tuning of the optical properties of the substituted variants of YAG was allowed by a relatively simple preparation of small cheap testing single crystals [21-27] or fibers [28-33] by the micro-pulling-down method (mPD). Nevertheless, the mPD, as well as the Czochralski method [34], are crucible methods; therefore both join similar experimental limitations.

Besides these classical crucible methods, the crucible-free floating zone method also has passed significant instrumental progress triggered by its utilization in the growth of single crystals of high-$T_c$ superconductors [35]. A very popular branch of the floating zone method (FZM) is its implementation to mirror optical furnaces with halogen or xenon lamps produced commercially as the so-called Travelling Solvent Zone Method (TSZM) [36-39]. Such an arrangement allows the melting and single crystal growth of materials with a melting point up to 3000°C based on oxides [15, 39-44], as well as a very recent series of metallic systems originally often unavailable by the Czochralski method [45-51].

The furnaces with laser heating are the most advanced instruments for single crystal growth by FZM [52]. The newly developed furnaces with multiple lasers based on modern laser diodes offers sharply demarcated hot zone with almost flat horizontal thermal gradient [53].

The usage of the floating zone method for the growth of the doped variants of YAG is limited because of the high mechanical strain resulting from the thermal gradient primarily in the vertical direction. The strain causes large cracks inside the single crystals. It is well-known problem, particularly in the case of Nd:YAG [54-56], nevertheless the narrow filaments were successfully grown by laser-heated pedestal growth technique [57].

It motivated us to develop and optimize the FZM implemented in a laser furnace to grow the single crystals of Nd:YAG as good reference material of the size of 4-6 mm diameter typical for flash tube cylinders used in many applications [11]. Such a production would reduce significantly the amount of waste material when the cylinders are drilled from the large ingots produced by the Czochralski method. Our developed technology is applicable also for single crystal growth of other classes of materials in an oxidizing environment generally unattainable by crucible Czochralski and mPD methods.

**EXPERIMENTAL**

Nd:YAG single crystals were grown from the powder precursor containing 0.6 % Nd mixed in Crytur Company in order to make a reliable comparison of the optical properties of our single crystals with those produced commercially by Czochralski method. The single crystal growth of Nd:YAG by the FZM was performed in two types of commercial optical furnaces. As a reference, a four-mirror furnace FZ-T-4000-VPM-PC from Crystal Systems Corporation, Japan, with four 1000W halogen lamps was used. The majority of the key experiments were performed in a laser diode furnace (LDF) FZ-LD-5-200W-VPO-PC-EG fabricated by an identical provider equipped with five 200 W laser diode units surrounding the hot zone in the horizontal plane. The diodes produce the near-infrared radiation of the wavelength $\lambda = 976 \pm 5$ nm of 4x8 mm$^2$ beam spot. A pyrometer is installed in LDF on a lift operating in the range $\pm 40$ mm below and above the hot zone. We used an empirical value of emissivity 0.870 calibrated by the melting point of Nd:YAG 1970°C. Commercial flash tubes produced by Crytur of length 30 mm and diameter 5 mm were used as seed single crystals. All growth processes were performed at Ar pressure 0.3 MPa of purity 6N.

The quality of the produced single crystals was verified by Laue method using both X-rays and neutrons; back-reflection geometry in the case of X-rays, CYCLOPS diffractometer at the Institut Laue-Langevin Grenoble in the case of neutrons [58]. The measured sharp nuclear diffractions were processed using Esmeralda software [59]. Specimens for optical characterization were prepared using a band saw Exact 300 CP with 100 microns thick band. The chemical analysis by PerkinElmer Lambda 950 instrument and the absorption spectra were recorded in Crytur Company.

## SINGLE CRYSTAL GROWTH PROCESS

### Precursor preparation

The polycrystalline precursor in the form of a rod is required for the single crystal growth by floating zone method. Its quality is crucial for the further stable and reproducible growth process. We have developed and tested several types of feeding and shaping steel forms of various inner diameters for the preparation of the precursor. The best results were achieved with the disassembled form of an inner diameter 8 mm displayed in Figure 1.

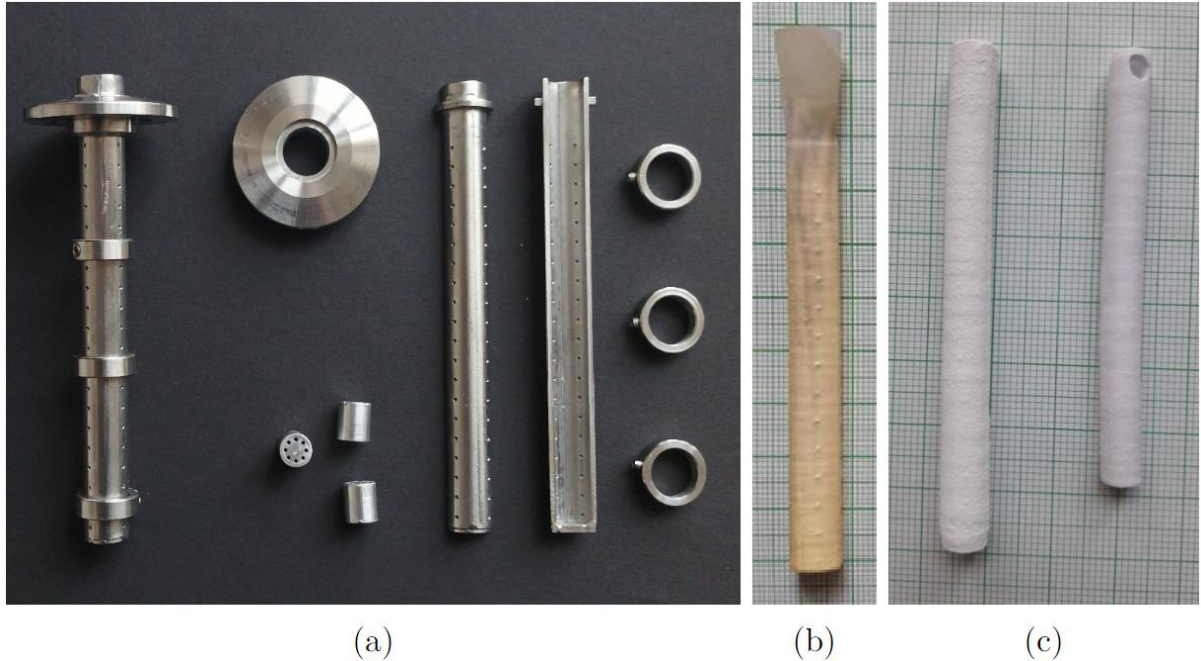

**Figure 1.** a) The feeding form for preparation of the polycrystalline precursor b) The latex capsule filled with powder precursor extracted from the form. c) The comparison of the volume of the rod before (left) and after (right) sintering process.

The assembled form was interposed into a vacuum chamber to stretch out the inner flexible latex capsule. The capsule was gradually filled with approximately 10 mm steps by powder precursor, regularly compressed by a piston since its fullness, and closed by a silicon plug on top. Then, the vacuum was unsealed, the steel form disassembled and the full latex capsule pulled out. In the next step, the latex form was cold pressed in a monostatic press by a pressure 3000 kg/cm$^2$. The latex was removed and the compressed precursor rod was sintered in the superkanthal furnaces protected by insertion into a sapphire tube. The sintering in air or vacuum was tested. The sintering under vacuum significantly improves the quality of the grown single crystals concerning microbubbles as well as the stability of the growth process as will be shown later. The precursor was always pre-sintered at 1000°C for 12 h, cooled down to drill a hole for the holder, and then sintered at 1600°C for 1 day on air or up to 5 days under the vacuum 10$^{-4}$mbar. The sintering process significantly increased the density which is reflected by the rod precursor shrinkage after the process, see Figure 1.

**Growth in mirror furnace**

In the first step, we tested the single crystal growth of Nd:YAG by FZM in a conventional four-mirror furnace equipped with halogen lamps as a reference. Even after several attempts, we were not able to achieve a stable process and a single crystal without cracks. We have identified two major reasons. The melting point of the Nd:YAG is close to the maximal attainable temperature in the given instrument. Only the weak violet color of the precursor does not enable it to absorb sufficient amout of the light. Therefore, the furnace was not able to keep a hot zone of sufficient volume even at 90-100% power. The height of the hot zone fluctuated significantly in the range 4 - 8 mm. The uncontrollable oscillations origin in significantly different absorption of melt, precursor rod, and grown single crystal. The melt is highly absorbing due to its honey-like color, while the single crystal is transparent. With the large volume of the melt in the hot zone, a greater amount of energy is absorbed, so the volume and size of the hot zone quickly expand and vice versa-see Figure 2. Such oscillations and particularly the moment of the avalanche growth of a single crystal induce enormous strain quickly released in the form of cracks. The regulation of the process to keep the constant volume of the melt was not reached. Various tested growth rates of 0.5 – 6 mm/h range did not influence the stability of the process.

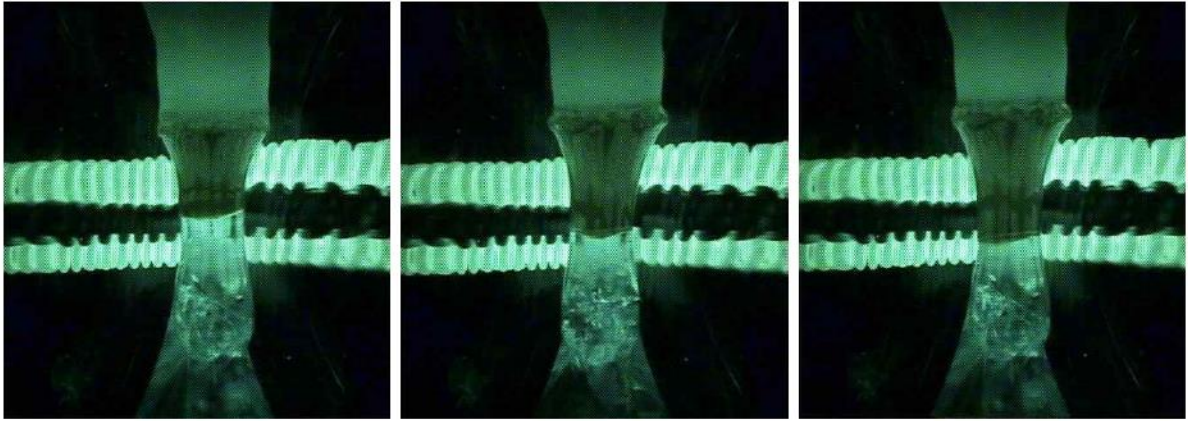

**Figure 2.** The detail of the hot zone in the mirror furnace recorded by the instrument camera at a constant power of the lamps. The dark well absorbing melt contrasts with the almost transparent single crystal below. The rapid growth of the melt volume is documented in the sequence of the panels from the left to right.

This oscillatory feedback resulting from a large difference in the light absorption of the melt and grown single crystal prevented us from performing successful growth of Nd:YAG in the halogen furnace.

**Growth in laser furnace**

A series of testing experiments were performed to optimize the parameters for a reproducible process of Nd:YAG single crystal growth in LDF. In the early growths, we were faced with similar problems as in the mirror furnace, however, to a much lesser extent signaling better-growing conditions, in general.

The cracks appear particularly only in the first 10 mm of the grown single crystal. The critical moment of the process is the connection between the seed crystal and precursor rod. While the melting point was already reached in the precursor rod, the transparent, only weakly absorbing seed crystal was significantly colder. Its sudden contact with the melt initiated instant cracking because of thermal shock. The cracks tend to propagate into a growing single crystal. Very simple solution is direct contact between the seed single crystal and the precursor rod during the ramping of the power up to the melting point not shorter than 6 hours. The heat is transferred into a seed single crystal which is sufficiently thermalized. The minor cracks which can occasionally appear are melted by the slow insertion of the small part of the seed rod into the hot zone.

The cracks did not appear in the further grown body of a single crystal in the conditions of stable parameters-constant pulling speed 1 mm/h. That is, properly centered both seed and precursor rod to avoid eccentric rotation, the constant diameter of the precursor rod, and particularly homogenous density of the material. However, even the crack-free single crystals conserved enormous frozen stress. This stress did not allow any machining and it was instantly released by cracking.

It motivated us to construct an afterheater which is commonly used in various designs in the Czochralski processes. The afterheater consists of the stage and central upper tube surrounding the single crystal just below the hot zone, both made of Pt/Rh alloy to be resistant to oxygen (Figure 3). The compact size allows the insertion of the afterheater inside the commercial quartz chamber produced by CSC Company floating zone furnace. The afterheater is fixed in the bottom on a screw with a low pitch for a precise location typically 2-3 mm below the laser beams. The operation principle is based on the back reflection of the IR radiation from the hot zone into the grown single crystal. The installation inside the laser furnace is possible thanks to the very precise focus of the laser beams; even at maximum power, the lasers cannot melt the material of the afterheater located 2-3 millimeters below the hot zone. The thermal profile inside the afterheater is estimated in Figure 4. The temperature of only 500°C was detected 20 mm below the hot zone in the case of the process without the afterheater. In contrast, a significantly higher temperature 1000°C was measured at the identical position just below the afterheater.

The usage of the afterheater requires very precise centering both of the seed single crystals and precursor rod to avoid their mutual contact. Moreover, the precise centering is one of the critical parameters responsible for the keeping constant diameter of the single crystals, which is essential for the reduction of the internal strain responsible for cracking.

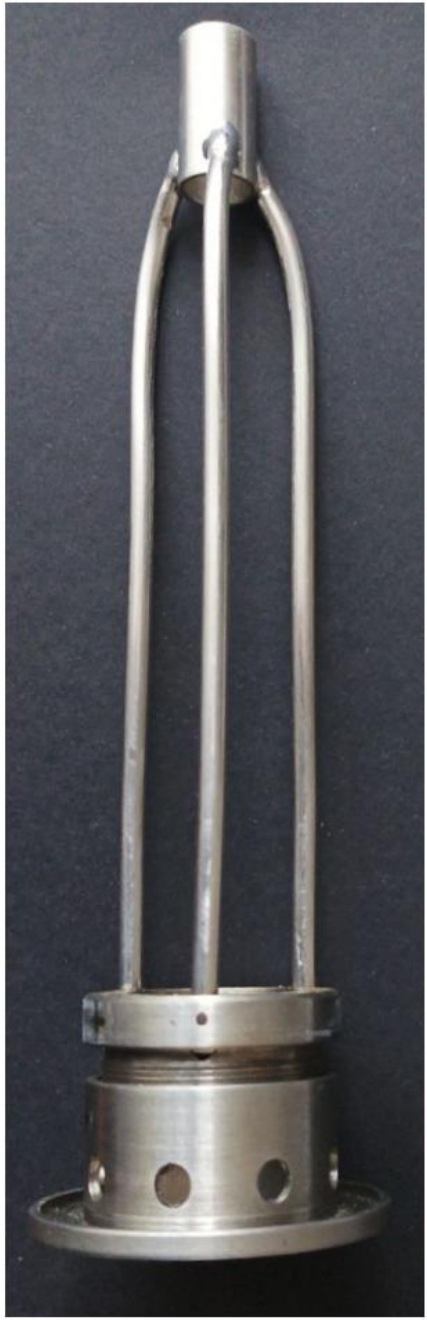
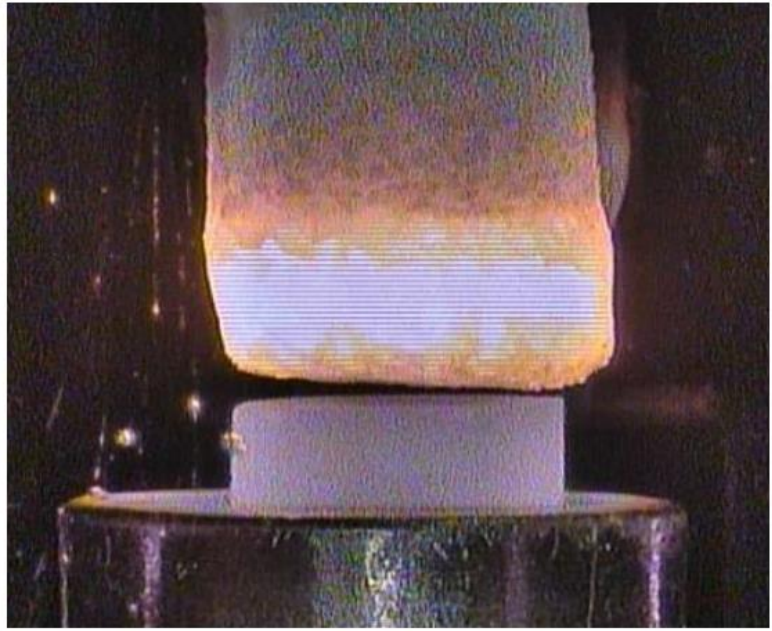
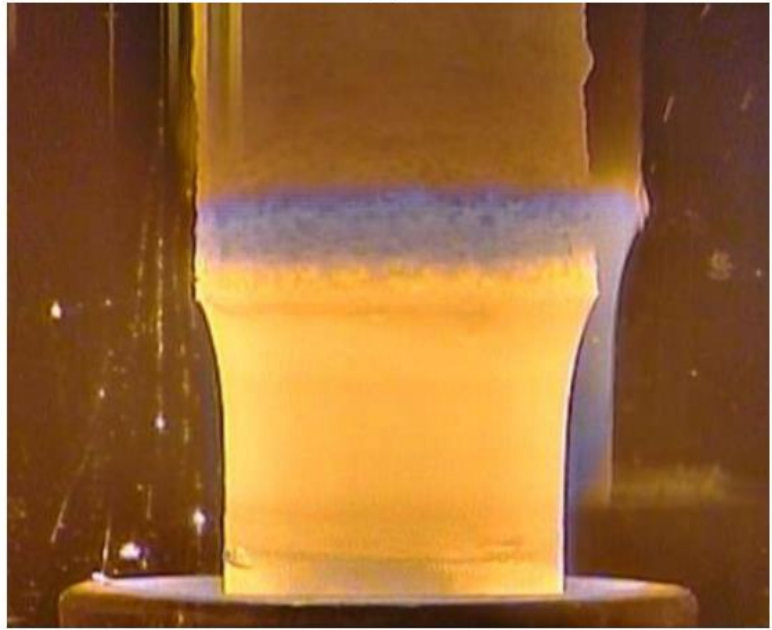

**Figure 3.** a) The design of the afterheater for the laser furnace, which fits the quartz chamber of the commercial laser furnace produced by CSC Company, b) the seed crystal inside afterheater during the ramping of the power of the lasers, c) the process of Nd:YAG single crystal grown with the installed afterheater.

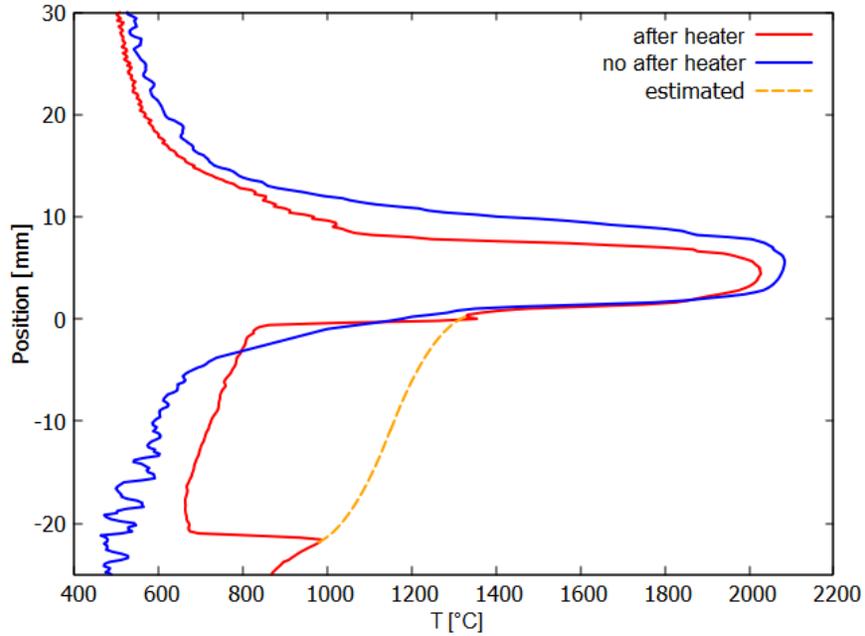

**Figure 4.** The in situ measured thermal profiles recorded by the movable thermometer. The temperature maximum corresponds to the relative position of the melt zone. The curves represent the temperature profile of the single crystal produced with (red) and without (blue) afterheater. The dashed line tentatively estimates the profile inside the afterheater. The drop in the temperature at the position of the afterheater is given by different emissivity of the metallic alloy.

The homogeneity and high density of the sintered material of the precursor rod is also crucial for the production of single crystals without bubbles. The high surface tension of the melt results in a very stable hot zone, on the other hand, very hardly allows the leakage of bubbles. Even a microcavity with a locked residual air can expand to a sizable bubble as a function of temperature which reaches ~2100°C inside the melt (Figure 5b). The spread microbubbles both incorporate into a single crystal or merge creating large bubbles which are usually able to overcome the surface tension and disappear. In the case of low-density precursor, the numerous small bubbles can create a large bubble that can collapse the hot zone when it bursts.

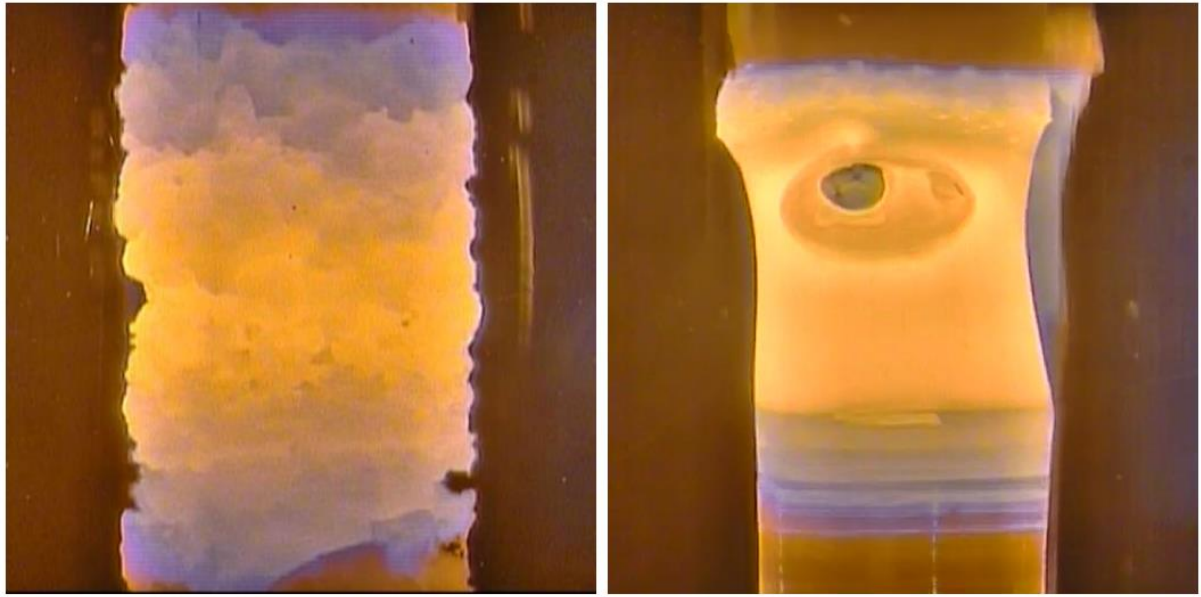
**Figure 5.** a) The process of the additional sintering of the precursor rod under a dynamic vacuum inside the chamber of the laser furnace. b) The macroscopic bubble inside the melt which had grown by merging of microbubbles coming from the low-density precursor rod.

The incorporation of the microbubbles into the single crystal is also very effectively reduced by suitable rotation of the seed and precursor rod responsible for the convection and vortexes inside the melt. The best results were reached with -5/20 rpm anticlockwise rotation (Figure 6).

We have also tested additional sintering directly in the laser furnace to reduce the residual gas inside the precursor rod. The power of the laser was set to heat the rod up ~1750°C under the dynamic vacuum. 1750°C is an unattainable temperature in the conventional process because of the possible reaction between the rod and sapphire protecting tube. We moved the rod at the highest possible speed of 200 mm/hour through the hot zone down and up (Figure 5a). The process leads to another contraction of the rod and any additional bubbles after this procedure were not detected inside the single crystal. However, the rough surface of the rod led to occasional imperfections in the diameter of the grown single crystal.

The optimal growing speed was found to be 1 mm/h; faster growth leads to higher inner strain and cracking of the single crystal. 1 mm/h is a compromise between the strain intensity and the time of the growth process.

**ANALYSIS OF THE SINGLE CRYSTAL QUALITY**

The basic characterization of the produced single crystals was performed visually-see Figure. 6 where four representative examples are shown demonstrating various stages of the optimization of the growth process. The "milky" single crystals containing microbubbles (Figure 6a) were produced in the case of low-quality polycrystalline precursor combined with the high rotation of both seed crystal and precursor rod -20/20 rpm. When the optimal rotation -5/20 was used with a significantly slower rotation of the precursor rod, the number of bubbles was substantially reduced: single crystal is transparent but the gas creates a certain amount of macroscopic bubbles often connected with the creation of the macroscopic bubble inside the melt (Figure 5b). The secondary effect of the bursting of the bubbles inside the melt is the irregular diameter of the single crystal along its length (Figure 6c). The improper centering of the seed rod is responsible for the bent single crystals (Figure 6b). The best single crystal is produced from high-quality precursor sintering under the vacuum and with the implemented afterheater (Figure. 6d). Single crystal is transparent, with constant violet color, bubbles-free, of almost constant diameter and with the reduced residual strain allowing its machining.

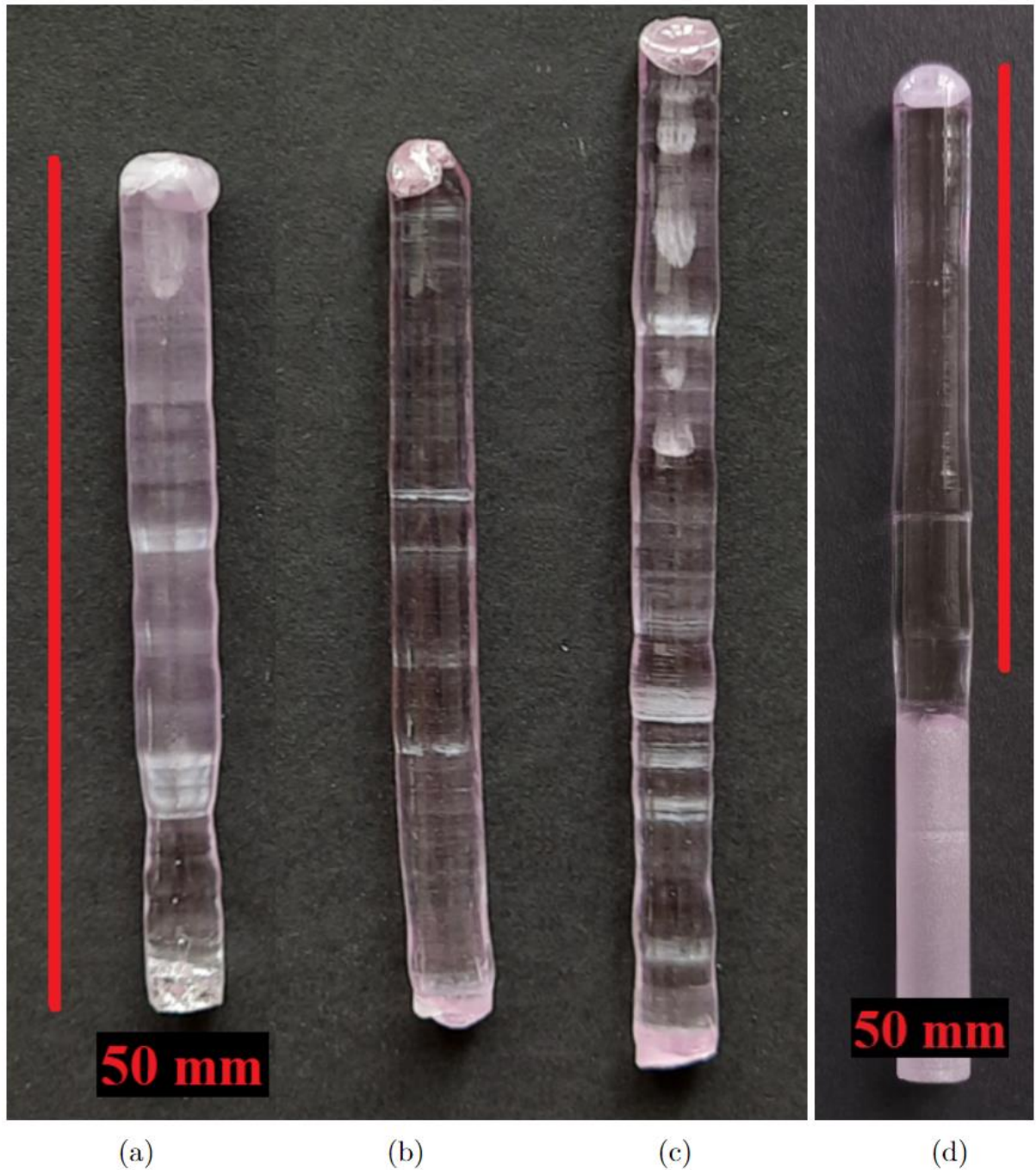

(a)  (b)  (c)  (d)

**Figure 6.** The Nd:YAG single crystals produced by FZM in a laser furnace. a) "milky" single crystal with microbubbles produced with rotation -20/20 rpm b) bent single crystals because of off-center seed rod c) irregularities in diameter caused by low-quality precursor rod containing gas cavities resulting in large bubbles in the melt d) high-quality single crystal produced with afterheater.

We inspected the single crystals under a microscope using laser dispersion. No microbubbles or dispersion centers were observed in crystals grown by optimized route. The crystallographic quality was tested by the X-ray Laue method, which has shown a sharp circular reflection signaling a high-quality material (Figure 7). We have evaluated the Laue pattern by OrientExpress software, which confirmed the cubic crystal structure of the reported lattice constant 12.01 Å [60].

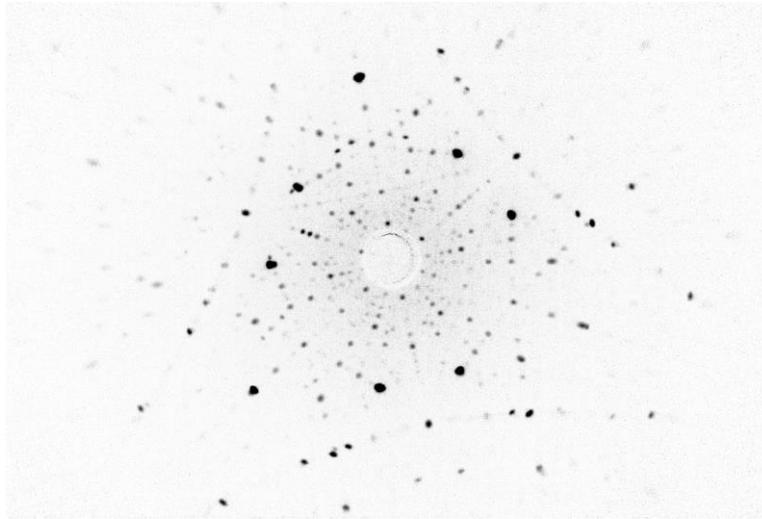

**Figure 7.** Representative Laue pattern (negative) of the produced Nd:YAG single crystal along [111] direction respecting the crystallographic orientation of the seed rod

We have employed neutron diffraction to quantify the stress and explain the stress frozen in single crystals produced without the afterheater [61]. In contrast to surface-sensitive X-ray Laue diffraction, the neutrons bring volume information about the lattice. A neutron diffraction pattern of the 0.3 cm$^3$ sample is displayed in Figure 8. Very sharp reflections confirm a high quality of the produced single crystal free of twinning and of low mosaicity. In the first step, the pattern was evaluated using model cubic structure (space group *Ia-3d*). The lattice parameter 12.080 Å was refined. However, displacement of a series of reflections from the expected positions was observed. Therefore, lattice parameters were refined independently in the next step. The model structure (space group *P1*) converged with parameters $a = 12.184$ Å, $b = 12.206$ Å, $c = 12.157$ Å, $\alpha = 90.022°$, $\beta = 90,029°$, $\gamma = 89,836°$, allowing the majority of the recorded reflections to be sufficiently described. Larger trigonal lattice parameters, compared to the cubic one, signalize a negative strain in the material. Moreover, all three principal crystallographic angles are deformed from the 90°. Any further annealing (1500°C for 4 weeks) did not reduce the strain to improve the machinability of the single crystals produced without after-heater.

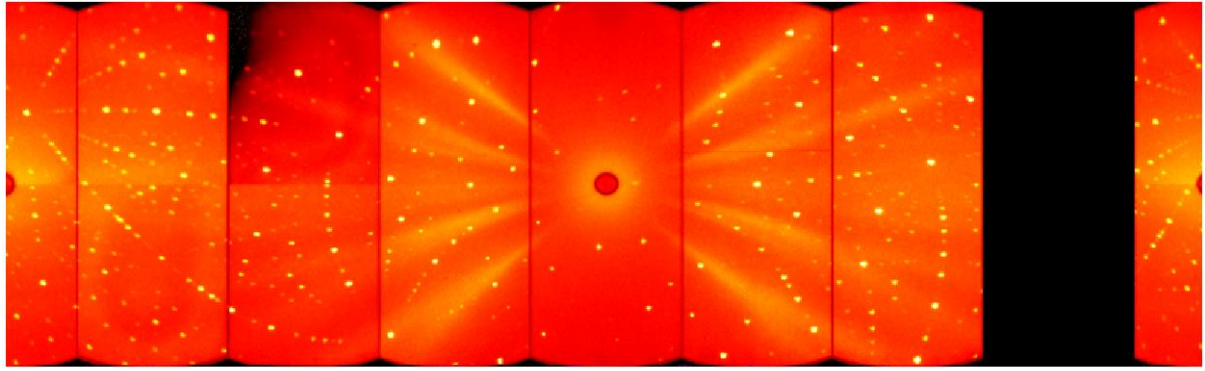
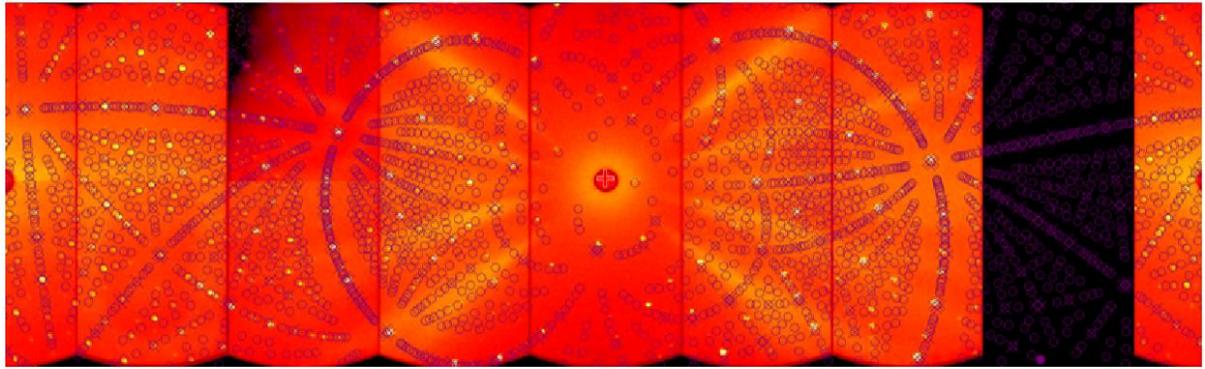

**Figure 8.** a) Neutron Laue pattern of the produced Nd:YAG single crystal, b) the pattern with the converged (*P1*) structural model.

Absorption spectrum in Figure. 9 is dominated by the *4f-5d* absorption of $Nd^{3+}$ below 245 nm (out of scale) and *4f-4f* transitions from the $^4I_{9/2}$ ground state to higher *4f* states represented by sharp peaks above 300 nm confirming the presence of the $Nd^{3+}$ ions in the matrices. Light scattering losses at both surfaces are responsible for non-zero values of absorbance out of the absorption peaks.

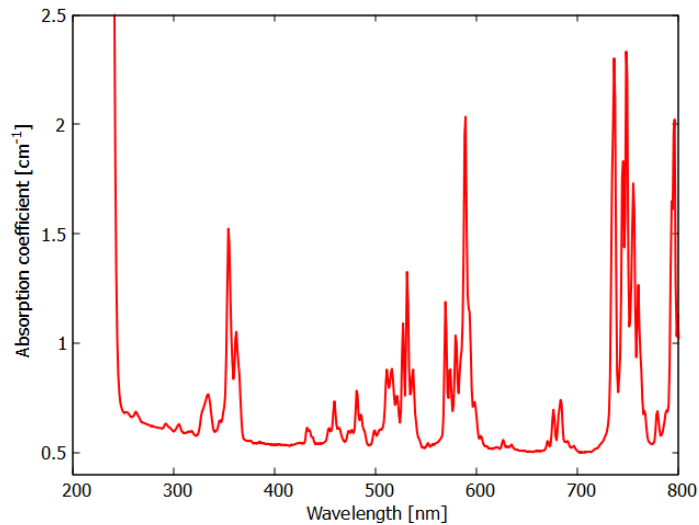

**Figure 9.** The absorption spectrum of the Nd:YAG single crystal produced with afterheater.

Due to the differences in ionic radii between $Y^{+3}$ and $Nd^{3+}$ ions, $Nd^{3+}$ ions are difficult to incorporate into the structure and the coefficient segregation is small, $K = 0.18$-$0.2$ [10, 62]. The concentration gradient in Nd:YAG single crystals is observed both in the vertical and horizontal directions within the slabs [62]. Our detailed analysis was performed on the surface of the middle part of a single crystal where the growing conditions are most stable. We have detected the concentration oscillating around 0.6 % $Nd^{3+}$ (Figure 10) which corresponds to the starting concentration of the powder precursor from Crytur company. Although the performed linear extrapolation is not ideally constant and a weak drop of Nd concentration is detected, segregation coefficient at given growing conditions is close to 1.

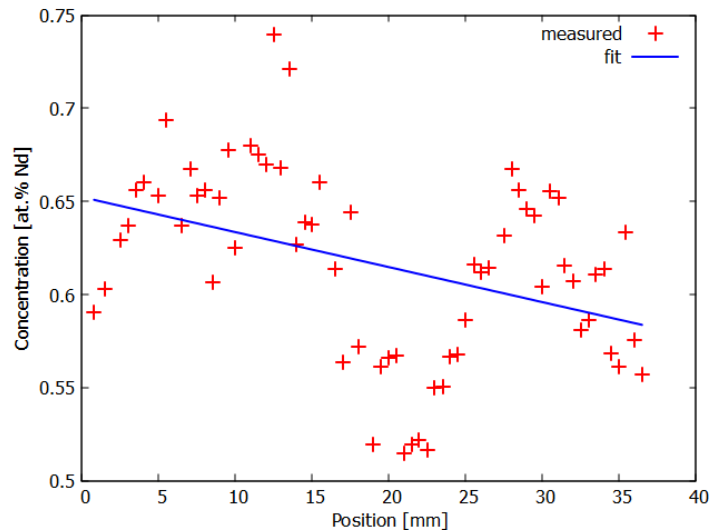

**Figure 10.** The analysis of the vertical concentration $Nd^{3+}$ gradient. The analysis was performed in Cryturu Company by PerkinElmer Lambda 950 instrument. The local minimum at position 20 mm is caused by an imperfection in single crystal diameter.

## CONCLUSIONS

We have successfully optimized the process of precursor preparation and single crystal growth of laser material Nd:YAG by the floating zone method implemented in the advanced laser furnace. The substantial reduction of the internal stress, which is a typical obstacle to producing single crystals by the floating zone method without the presence of cracks, was achieved by a high-quality precursor rod and particularly by the installation of the metallic afterheater. The afterheater reflects the IR radiation and thermalizes the growing single crystal. The precise focus of the laser beams allows the installation of the afterheater just in the vicinity of the melting zone, which is not possible in common mirror furnaces. The analyzed quality of the produced single crystals is comparable with those produced for commercial applications by Czochralski method. We have already successfully tested the developed technology for the production of the Ce:YAG and Ce:GGAG single crystals under the pressure of pure oxygen. Results are to be published soon. It confirms the wide applicability of the method to grow single crystals of oxide-based materials of high melting points unattainable in crucible methods, moreover in an oxidizing environment which improves their expected optical properties.

## AUTHOR INFORMATION

**Corresponding Author**

**Jiří Pospíšil** – *Charles University, Faculty of Mathematics and Physics, Department of Condensed Matter Physics, Ke Karlovu 5, 121 16 Prague 2, Czech Republic*

**Authors**

**František Zajíc** – *Charles University, Faculty of Mathematics and Physics, Department of Condensed Matter Physics, Ke Karlovu 5, 121 16 Prague 2, Czech Republic*

**Martin Klejch** – *CRYTUR, Na Lukách 2283, 511 01 Turnov, Czech Republic*

**Adam Eliáš** – *Charles University, Faculty of Mathematics and Physics, Department of Condensed Matter Physics, Ke Karlovu 5, 121 16 Prague 2, Czech Republic*

**Milan Klicpera** – *Charles University, Faculty of Mathematics and Physics, Department of Condensed Matter Physics, Ke Karlovu 5, 121 16 Prague 2, Czech Republic*

**Alena Beitlerová** – *Institute of Physics, Academy of Sciences of the Czech Republic, Cukrovarnicka 10, 162 53 Prague, Czech Republic*

**Martin Nikl** – *Institute of Physics, Academy of Sciences of the Czech Republic, Cukrovarnicka 10, 162 53 Prague, Czech Republic*

## ACKNOWLEDGEMENT

This work is part of the research program GAČR 21-17731S which is financed by the Czech Science Foundation. The single crystals have been grown by FZM in the Materials Growth and Measurement Laboratory MGML (MGML.eu) which is supported within the program of Czech Research Infrastructures (Project No. LM2018096). We also thank the technical and material support from the Crytur Company and also the staff A. Balášová for precursor preparation and M. Martínek and Š. Uxa cooperating on analysis of the single crystal quality. The authors are also indebted to Dr. Ross H. Colman for critical reading and correcting of the manuscript.